# POINT DEFECTS INFLUENCE ON THE RAMAN SPECTRUM OF THE NAPHTHALENE

## M. A. Korshunov

*LV Kirensky Institute of Physics, Siberian Branch of Russian Academy of Science, Krasnoyarsk, 660036 Russia*
*E-mail:* mkor@iph.krasn.ru

**Abstract:** The carried out polarization examinations of $C_{10}H_8$ spectrums of small frequencies of a Raman effect of light has revealed the presence of additional lines of small intensity. Calculation of histograms of spectrums of the lattice oscillations is carried out by Dyne's method. It is shown, that the presence of vacancies in structure of $C_{10}H_8$ leads to occurrence of additional lines around 90 cm$^{-1}$. Occurrence of lines at frequencies less then 30 cm$^{-1}$ is attributed to the presence of orientation disorder of molecules. Apparently, structure $C_{10}H_8$ has both types of defects, both orientation disorder of molecules and vacancies, which causes occurrence of additional lines in a spectrum of small frequencies.

Research techniques of spectral properties of ideal molecular crystals developed well enough. Actual crystals have various defects which presence affects crystal's physical properties. At present time, similar organic crystals of low symmetry find the increasing application in molecular electronics. Perceptivity of use of similar crystals for record and information processing is remarked. At the same time, one of the restrictions, interfering diminution of the sizes of the information record area, is bound to the presence of diffusion in a sample. Presence of diffusion can be caused by presence of vacancies. Before studying migration of molecules it is necessary to determine presence of vacancies in the given substance. A perspective method that determines the presence of vacancies in substance, can be Raman scattering of light with of small frequencies, because the presence of vacancies in a crystal affects the lattice oscillations and will be observed in spectrums [1]. Thus it is better to carry out such examination on objects well investigated by various methods.

Table. Frequencies of orientation oscillations and their symmetries for $C_{10}H_8$.

| Frequency, cm$^{-1}$ | 46.0 | 51.5 | 71.5 | 73.5 | 107.5 | 123.5 |
|---|---|---|---|---|---|---|
| Type | $B_g$ | $A_g$ | $B_g$ | $A_g$ | $A_g$ | $B_g$ |

In the present work calculations of spectrums of the lattice oscillations at presence of vacancies in the structure are carried out and comparisons with the experimental spectrums obtained by polarization examinations is done. Previously, using the same approach for parasubstituted benzole [2] it has been shown, that the presence of vacancies results in occurrence of additional lines of small intensity near of 70 cm$^{-1}$. In this paper analogous examinations for naphthalene crystals ($C_{10}H_8$) are given.

According to X-ray diffraction data [3] $C_{10}H_8$ crystallizes in centrosymmetric space group P2$_1$/a with two molecules in a unit cell. In the calculations of frequency spectrums the molecular structure is supposed to be absolutely rigid. Interaction between molecules was described using method of atom-atom potentials [4]. Calculations were carried out with use of Dyne's method [5]. This method allows to carry out calculations of frequency spectrums of disordered molecular crystals. As a result, the histograms that show probability of appearance of spectral lines in the picked frequency interval have been obtained.

Spectrums of the lattice oscillations $C_{10}H_8$ were studied by various authors [6, 7]. In these papers the six intensive lines of orientational oscillations of molecules were found in the spectrum, but additional lines of small intensity were not observed. According to IR absorption data [8], translational oscillations have values 66.0 and 96.0 cm$^{-1}$.

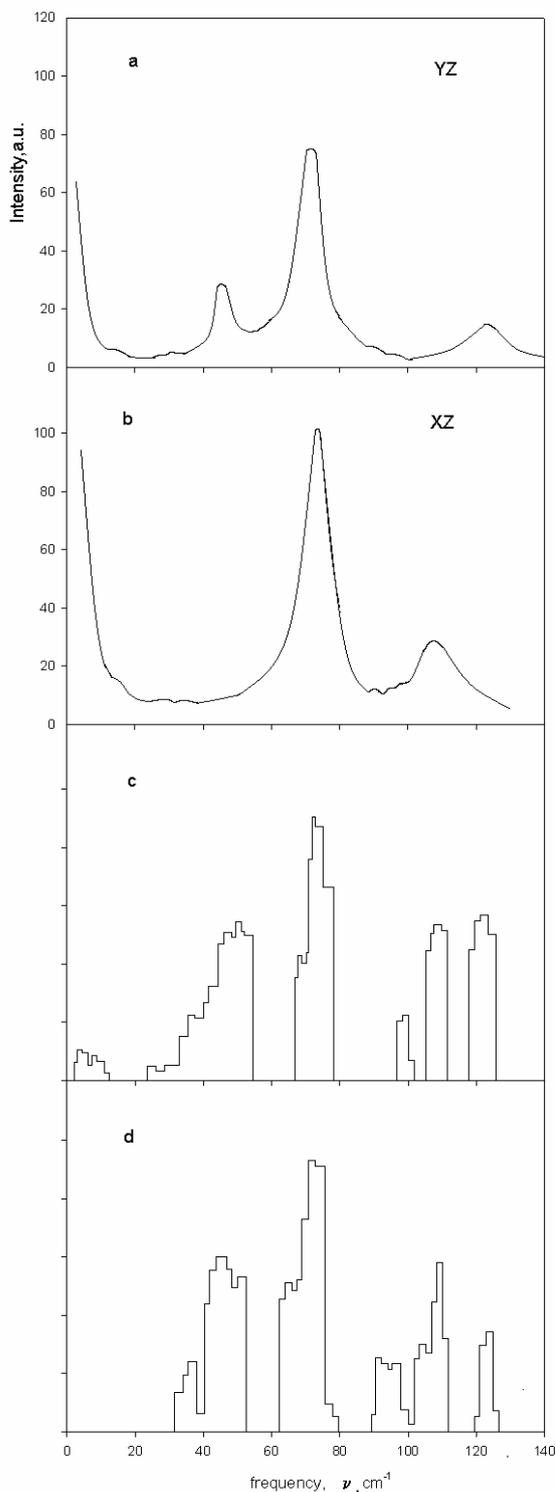

Figure 1.

Polarization examinations of spectrums $C_{10}H_8$ at temperature 290 K, carried out in the present paper, also shown that in a spectrum there are six intensive lines and a series of lines of small intensity are observed. In figures 1a and 1b components (XZ) and (YZ) of the polarized spectrums of $C_{10}H_8$ are presented. Frequencies of intensive lines are presented in the Table. Also a series of additional lines of small intensity 14.5, 27.5, 31.0, 90.0, and 96.0 cm$^{-1}$ is observed.

For an explanation of occurrence of additional lines in a spectrum of the lattice oscillations of $C_{10}H_8$, calculations are carried out. Occurrence of these lines can be determined by the presence of orientation disorder of molecules or by the presence of vacancies.

In figure 1c the histogram of spectrum $C_{10}H_8$ obtained by Dyne's method with orientation disorder taken into account, but without vacancies. It is clear, that additional lines in area below 30 cm$^{-1}$ appeared and some lines are broadened, but there are no lines in the area of 90.0 cm$^{-1}$.

Beside the orientation disorder the vacancies could be present in the crystal structure. Results of calculations for this case are presented in figure 1d. A one can see from the figure, on a histogram there are additional lines about 90.0 cm$^{-1}$, and broadening of some lines, but lines in the area lower then 25 cm$^{-1}$ are not present.

Thus we assume, that the presence of vacancies in the structure of $C_{10}H_8$ affects occurrence of additional lines in the field of 90 cm$^{-1}$. Occurrence of lines in the area lower then 30 cm$^{-1}$ is determined by the presence of orientation disorder of molecules. For both cases broadening of the lines corresponding to orientation oscillations is found. Hence, the structure of $C_{10}H_8$ has both types of defects, both orientation disorder of molecules and vacancies, which causes occurrence of additional lines in a spectrum of small frequencies.